\documentclass[fleqn,twoside]{article}
\usepackage{espcrc2}

\usepackage{graphicx}
\usepackage{epsfig}
\usepackage[figuresright]{rotating}

\usepackage{xspace}

\newcommand{\AmS}{{\protect\the\textfont2
  A\kern-.1667em\lower.5ex\hbox{M}\kern-.125emS}}

\def\cref#1{Chapt.\,\ref{#1}}
\def\Cref#1{Chapter~\ref{#1}}
\def\sref#1{Sect.\,\ref{#1}}

\def\fref#1{Fig.\,\ref{#1}}

\def\rref#1{Ref.\,\cite{#1}}

\def\1{\footnotemark[1]}
\def\and{\& }
\def\Cerenkov{\v{C}erenkov\xspace}

\def\deg{$^\circ$\xspace}
\def\etal{et al.\xspace}
\def\gcm2{g/cm$^2$\xspace}

\def\Xmax{$X_{max}$\xspace}

\def\Caption#1{\vspace*{-9mm}\caption{#1\vspace{-4mm}}}

\title{Experimental Efforts on Very High-Energy Cosmic Rays and their
       Interactions --- Conference Summary}

\author{
J.R.~H\"orandel\address{Radboud University Nijmegen, Department of
Astrophysics, Nijmegen, The Netherlands; j.horandel@astro.ru.nl}
\vspace*{-0.01cm} 
}

\begin{document}

\begin{abstract}
Progress reported during the XV International Symposium on Very High-Energy
Cosmic-Ray Interactions is summarized. Emphasize is given to experimental work.
The actual status, recent results, and their implications on the present
understanding of the origin of high-energy cosmic-rays and their interactions
are discussed.
\vspace{1pc}
\end{abstract}

\maketitle

\section{Introduction}
\begin{figure}
 \epsfig{file=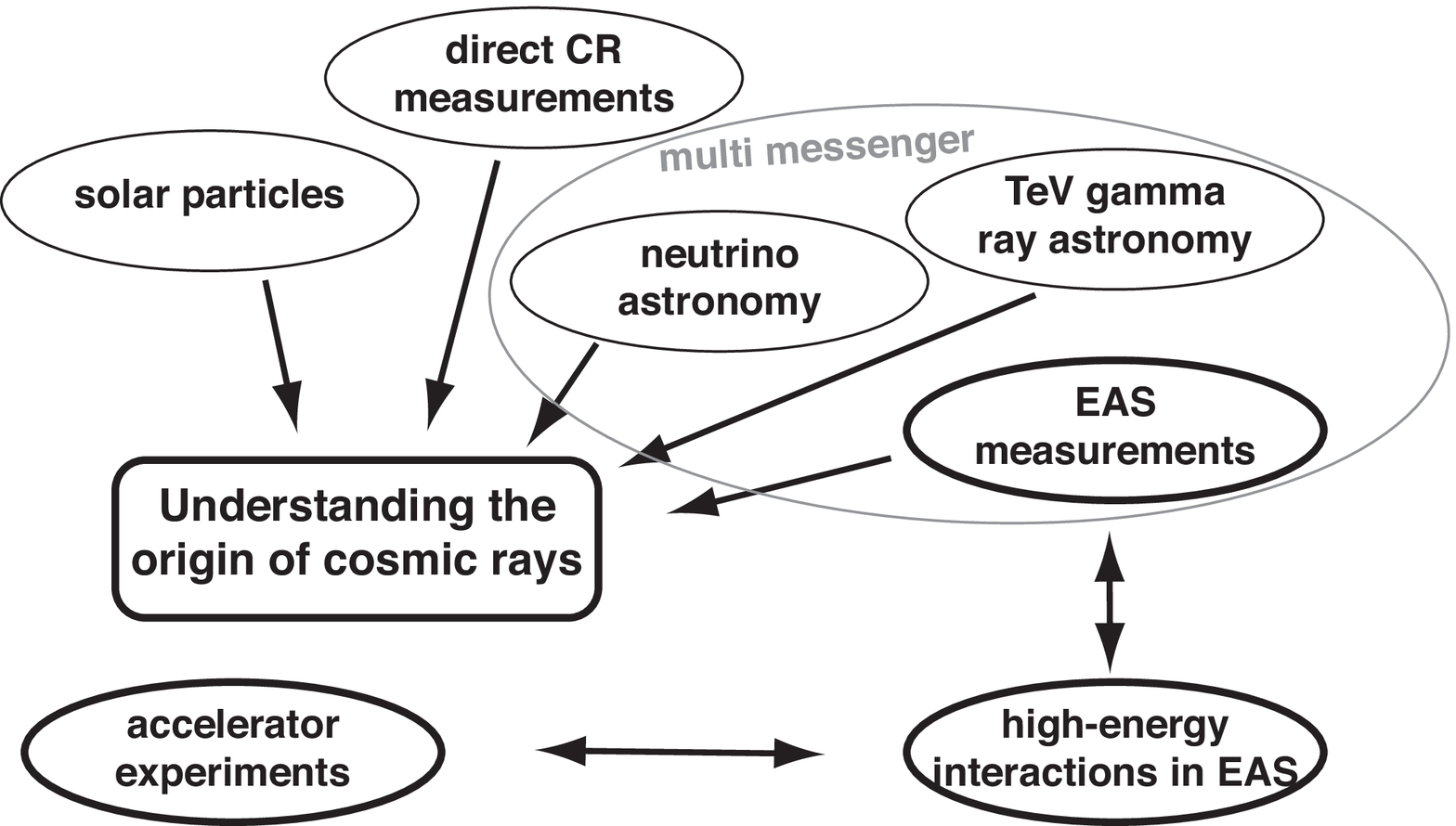, width=\columnwidth}
 \Caption{Schematic view of the relations between the topics discussed at the
   meeting.}
 \label{interplay}
\end{figure}

The XV International Symposium on Very High-Energy Cosmic-Ray Interactions
(ISVHECRI 2008) took place in Paris from September $1^{st}$ to $6^{th}$, 2008.
Main objective of the symposium was to contribute to the understanding of the
origin of cosmic rays. The relation between the various topics discussed is
shown schematically in \fref{interplay}.  Progress has been reported in the
fields of solar particles (\sref{solsec}), TeV gamma-ray astronomy
(\sref{gammasec}), neutrino astronomy (\sref{neutrinosec}), as well as the
direct (\sref{directsec}) and indirect (\sref{eassec}) measurements of cosmic
rays.  To reveal the origin of high-energy cosmic rays combined efforts of
TeV gamma-ray astronomy, neutrino astronomy, and cosmic-ray research will
become crucial (multi messenger approach).  The properties of hadronic
interactions are studied in detail at accelerator experiments
(\sref{accelsec}). This knowledge contributes to the understanding of
high-energy interactions in air showers. Also information deduced from air
showers can be used to deduced properties of high-energy interactions in turn
(\sref{wwsec}).  This is in particular of interest in kinematical and energy
regions not covered by accelerator experiments.  The (astrophysical)
interpretation of air shower data depends critically on our understanding of
the properties of high-energy interactions.  Thus, accelerator experiments
contribute significantly to our understanding of the origin of cosmic rays.

The ISVHECRI meeting series has served and should continue to serve in future
as bridge between the accelerator and cosmic-ray communities. The relations
between this fields should be further strengthened. Thus, improving our
knowledge on high-energy interactions on one side, but at the same time also
increasing our (astrophysical) understanding of the processes in the Universe.

This article summarizes experimental work presented at the meeting.
\footnote{Work reported, related to emulsion chambers and theoretical
considerations are discussed in separate summaries. Nevertheless, more than
20~h of talks and some 30 posters are expected to be covered by the present
article. This is an almost impossible venture, thus, this summary may be
biased towards a personal selection of scientific highlights.} The actual status,
recent results, and their implications on the present understanding of the
origin of very high-energy cosmic rays and their interactions are discussed.

\section{High-Energy Cosmic Rays and Extensive Air Showers}
The Earth is permanently exposed to a vast flux of high-energy particles from
outer space. Most of these particles are ionized atomic nuclei with
relativistic energies.  They have a threefold origin.  Particles with energies
below 100~MeV originate from the Sun \cite{ryanpune,kahlerpune}.  Cosmic rays
in narrower sense are particles with energies from the 100~MeV domain up to
energies beyond $10^{20}$~eV. Up to several 10~GeV the flux of the particles
observed at Earth is modulated on different time scales by heliospheric
magnetic fields \cite{fichtnerpune,heberecrs}.  Cosmic rays with energies below
$10^{17}$ to $10^{18}$~eV are usually considered to be of galactic origin
\cite{gaisserstanev,gaisserjapan,smparnp,pg,cospar06,ricap07,behreview}.
Particles at higher energies can not be magnetically bound to the Galaxy.
Hence, they are considered to be of extra-galactic origin
\cite{behreview,naganowatson,bergmanbelz,kamperttaup07,wuerzburg}. 

When high-energy cosmic rays impinge on the atmosphere of the Earth they
initiate cascades of secondary particles --- the extensive air showers.
Objective of air shower detectors is to derive information about the shower
inducing primary particle from the registered secondary particles.  Addressing
astrophysical questions with air-shower data necessitates the understanding of
high-energy interactions in the atmosphere.  

For air shower interpretation the understanding of multi-particle production in
hadronic interactions with small momentum transfer is essential
\cite{engelpylos}.  
Due to the energy dependence of the coupling constant $\alpha_s$ soft
interactions cannot be calculated within QCD using perturbation theory.
Instead, phenomenological approaches have been introduced in different models.
These models are the main source of uncertainties in simulation codes to
calculate the development of extensive air showers, such as the program CORSIKA
\cite{corsika}.  Several codes to describe hadronic interactions at low
energies ($E<200$~GeV; e.g.\ GHEISHA \cite{gheisha} and FLUKA
\cite{flukacern,flukaCHEN}) as well as at high energies (e.g.\ DPMJET
\cite{dpmjet}, QGSJET \cite{qgsjet,qgsII,qgsjetII}, SIBYLL \cite{sibyll21}, and
EPOS \cite{epos,epos2}) have been embedded in CORSIKA.

\section{Accelerator Experiments}\label{accelsec}

\subsection{Accelerator data needed for cosmic-ray physics}

More information about hadronic interactions is needed from accelerator
experiments to fully understand cosmic rays, as discussed in the following.

\paragraph{Air shower measurements}
In high-energy interactions most energy is escaping the interaction region in
the forward direction, i.e.\ at large pseudo rapidity values \footnote{The
pseudo rapidity $\eta$ describes the angle of a particle relative to the beam
axis. Is is defined as $\eta=-\ln\left[\tan\left(\theta/2\right)\right]$, where
$\theta$ is the angle between the particle momentum $\vec{p}$ and the beam
axis, or, using the longitudinal component $p_L$ of the particle momentum
$\eta=1/2\ln\left[(|\vec{p}|+p_L)/(|\vec{p}|-p_L)\right]$.} $\eta$. For
example, the energy flow at the LHC at $E_{cm}=14$~TeV, corresponding to
$E_{lab}=E_{cm}^2/m_p=10^{17}$~eV, peaks at pseudo rapidity values around 7 to
10. The forward region with values $|\eta|>4$ is of great importance for air
shower experiments.

Of particular interest are the total (inelastic) cross sections, the
elasticity/inelasticity of the interactions, as well as the production cross
sections of secondary particles and their parameter distributions, like
multiplicity, transverse momentum, energy, and pseudorapidity.  As projectiles
protons and pions are of interest to study the elementary interactions but also
beams of heavier nuclei (such as C, N, O, or Fe, being dominant in the primary
cosmic-ray composition) are desirable.  Targets are preferably air
constituents, i.e.\ nitrogen, oxygen, (and carbon). 
In particular, at the LHC the study of p-p  and p-N interactions is of
great importance.

\begin{figure}
 \epsfig{file=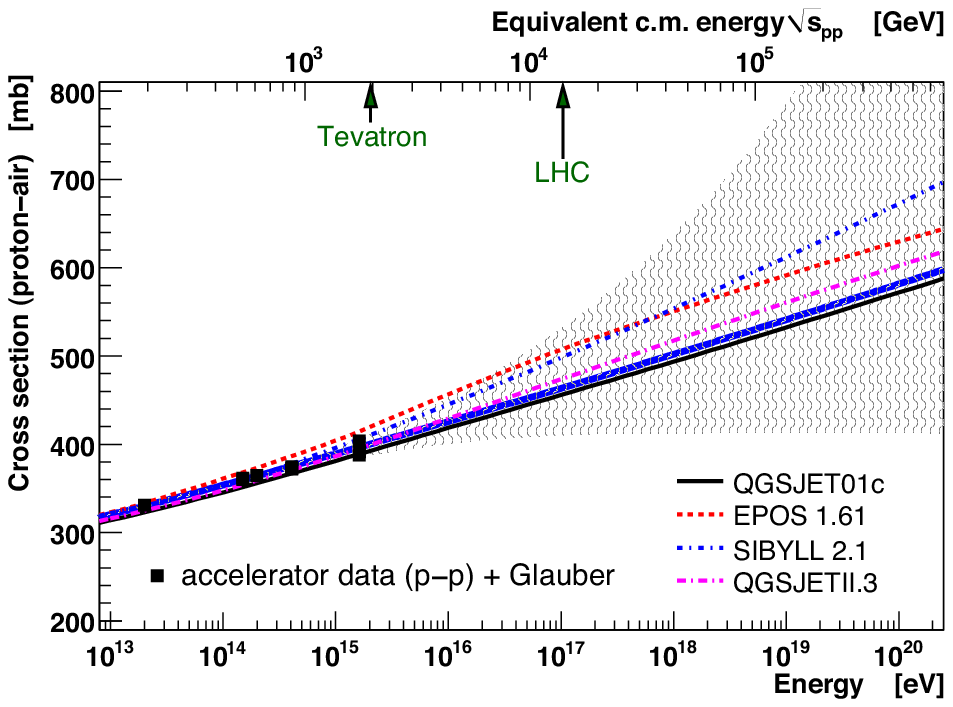, width=\columnwidth}
 \Caption{Uncertainties of the extrapolation of the proton-air cross section
          from accelerator to cosmic-ray energies \cite{Ulrich:2009yq}.}
 \label{ppextrap}
\end{figure}

The uncertainties introduced in the proton air cross section by extrapolating
from accelerator data to highest energies is illustrated in \fref{ppextrap}
\cite{Ulrich:2009yq}. It is obvious that LHC data will drastically reduce the
uncertainties in the regime of the highest-energy cosmic rays.

\paragraph{Direct measurements} 
Further input from accelerator experiments is also required for the
interpretation of data from balloon borne cosmic-ray detectors. The systematic
uncertainties of measurements of the boron-to-carbon ratio, see \fref{creamBC},
are presently dominated by uncertainties in the production cross section of
boron in the residual atmosphere above the detector. Boron is produced through
spallation of the relatively abundant elements of the CNO group in the
atmosphere.
\footnote{The detectors float typically below a residual atmosphere of about
$3-5$~\gcm2.}
Thus, the production cross sections of boron for protons and CNO nuclei
impinging on nitrogen targets are of great interest at energies significantly
exceeding 100~GeV/n.

\subsection{Present activities}

\begin{figure}
 \epsfig{file=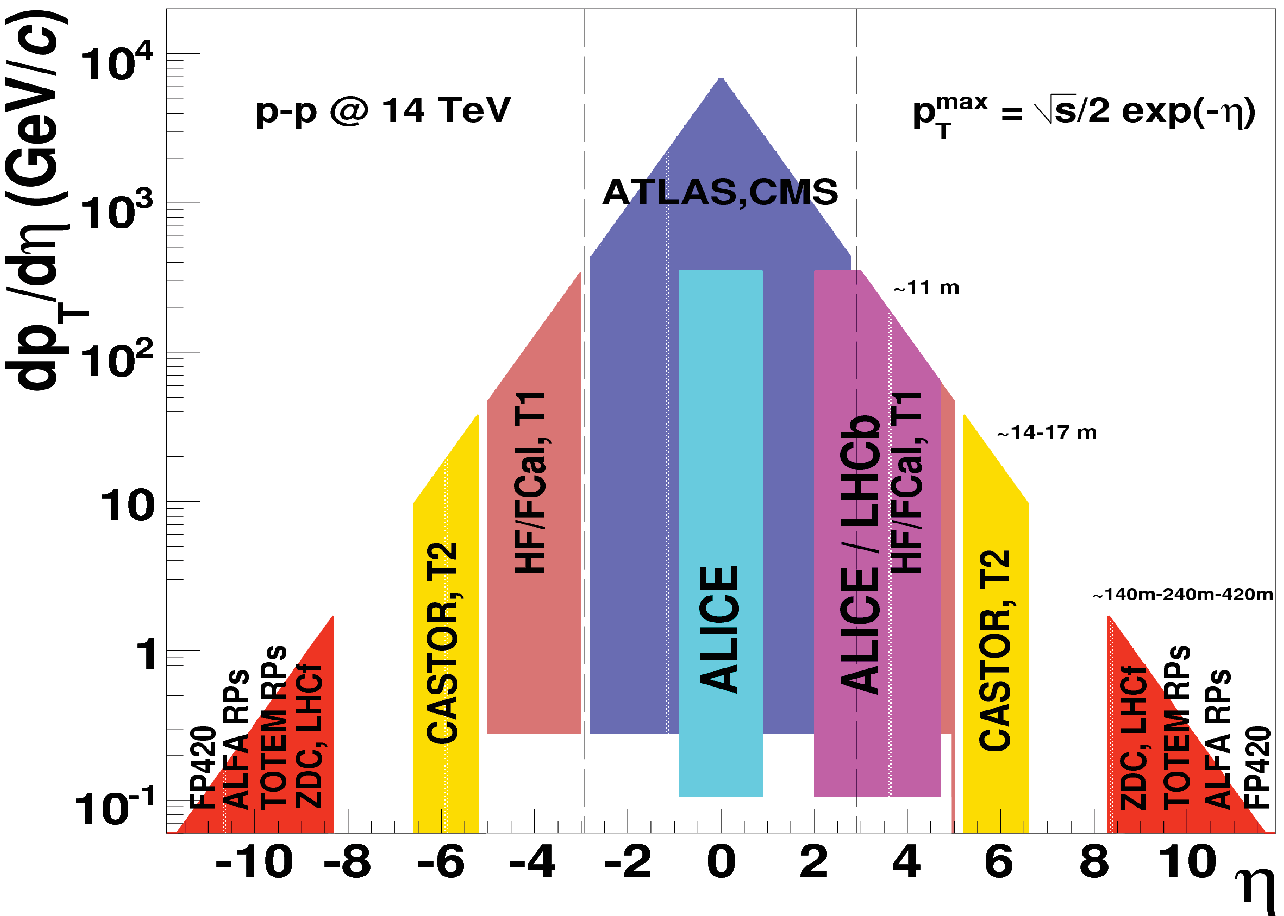, width=\columnwidth}
 \Caption{Pseudorapidity ranges covered by various experiments at the LHC
          \cite{McCauley}.}
 \label{LHCpseudorap}
\end{figure}

The (re)start of the LHC \cite{lhc} during 2009 will begin a new era in
particle physics.  The general purpose detectors ATLAS \cite{atlas,dova} and
CMS \cite{cms,McCauley} as well as the specialized ALICE \cite{alice} and LHCb
\cite{lhcb} experiments cover the central collision region with $|\eta|<4$, see
\fref{LHCpseudorap}.  The most important experiments for air shower physics are
experiments exploring the forward region $|\eta|>4$, such as LHCf \cite{lhcf}
and TOTEM \cite{totem}. The principle idea is to measure charged particles as
close as possible to the beam, sometimes even with detectors inside the beam
pipe and to register neutral particles in forward direction (zero degree
detectors).

LHCf \cite{tricomi,lhcf} comprises two sampling calorimeters (tungsten absorber
and scintillator layers), located $\pm140$~m from the ATLAS interaction point
and covers pseudo rapidities $|\eta|>8.5$. Two imaging shower
calorimeters are located on the beam axis. One detector uses scintillating
fibers and multi-anode PMTs for the read-out, the second calorimeter uses
silicon strip sensors.  In addition, ATLAS \cite{atlas} comprises several
forward detector systems, namely: LUCID ($5.4<|\eta|<6.1$), ALFA, and a zero
degree calorimeter (ZDC, $|\eta|>8.3$).

The TOTEM experiment \cite{eggert,totem} is dedicated to the measurement of the
total proton-proton cross section. The detectors are located on both sides of
the interaction point close to the CMS experiment and cover the pseudo rapidity
range $3.1\le|\eta|\le6.5$. Two telescopes for inelastically produced charged
particles are installed 9~m and 13.5~m from the interaction point,
respectively.  Two so-called "Roman Pots" are placed 147~m and 220~m from the
interaction point, designed to detect leading protons at merely a few mm from
the beam center.
The CMS experiment \cite{cms} comprises also forward detectors: CASTOR
($5.1<|\eta|<6.6$) and a ZDC ($|\eta|>8.3$).
New detectors for the extreme forward region are planned for both, ATLAS and
CMS at 220~m and 420~m distance from the interaction points.

Also important for air shower physics, are studies of the properties of
hadronic interactions at fixed target experiments \cite{panman}. The particle
beams for those experiments are extracted from the main accelerator rings,
thus, reaching limited energies only. Some of the most important present
experiments are BNL-E910, HARP, MIPP, and NA49/NA61.

E910 was a fixed-target proton-nucleus (p-A) experiment performed at the
Brookhaven Alternating Gradient Synchrotron (AGS) reaching momenta up to
18~GeV/c \cite{E910,Chemakin:1999jd,Chemakin:2007nb}.

HARP is a  large solid angle experiment to measure hadron production using
proton and pion beams with momenta between 1.5 and 15 GeV/c impinging on many
different solid and liquid targets from low to high Z (H-Pb) \cite{harp}. The
experiment, located at the CERN PS, took data in 2001 and 2002.  For the
measurement of momenta of produced particles and for the identification of
particle types, the experiment includes a large-angle spectrometer, based on a
Time Projection Chamber and a system of Resistive Plate Chambers, and a forward
spectrometer equipped with a set of large drift chambers, a threshold \Cerenkov
detector, a time-of-flight wall and an electromagnetic calorimeter.  The large
angle system uses a solenoidal magnet, while the forward spectrometer is based
on a dipole magnet. 

The Main Injector Particle Production (MIPP) experiment is operated in the FNAL
Meson Line \cite{mipp,mipp2,sorel}.  MIPP targets were exposed to 120 GeV/c
protons from the Main Injector, and to lower-energy secondary beams of mixed
composition. The MIPP detector configuration is fairly similar to the HARP one,
with some differences due to the higher momenta probed. Drift chambers are used
to track beam particles, and a beam threshold \Cerenkov detector is used to tag
the beam particle type. Tracks are reconstructed making use of two dipole
magnets deflecting secondary particles in opposite directions, plus a TPC,
drift chambers, and proportional wire chambers. As for HARP, the TPC, a
time-of-flight wall (ToF), and a threshold \Cerenkov detector are used for
particle identification. In addition, MIPP uses also a Ring Imaging \Cerenkov
detector (RICH) for high-momentum particles.

The NA49 detector is a wide acceptance spectrometer for the study of hadron
production in proton-proton, proton-nucleus, and nucleus-nucleus
collisions at the CERN SPS \cite{NA49}. The main components are 4 large-volume
TPCs for tracking and particle identification via dE/dx. Time-of-flight
scintillator arrays complement particle identification. Calorimeters for
transverse energy determination and triggering, a detector for centrality
selection in p+A collisions, and beam definition detectors complete the set-up.
The experiment has been upgraded to NA61 (SHINE -- SPS Heavy Ion and Neutrino
Experiment) \cite{shine}.  The experimental program till 2013 foresees
comprehensive studies of hadron production in proton-proton, proton-nucleus,
and nucleus-nucleus interactions, including proton-lead interactions.
Also runs with secondary beams ($\pi$, K, p) with momenta between 30 and
350~GeV/c are planned, including measurements with a carbon target.

These experiments have offered/will offer large data samples useful for the
cosmic-ray community. Data from E910, NA49, and HARP have been published
\cite{panman}. Results from MIPP and NA61 are expected in the near future.

The experiments BRAHMS \cite{brahms}, PHENIX \cite{phenix}, PHOBOS
\cite{phobos}, and STAR \cite{star} at the Relativistic Heavy Ion Collider
(RHIC) at Brookhaven National Laboratory study mainly collisions of heavy
nuclei \cite{lappi}.  Of interest for the cosmic-ray community are interactions
of gold nuclei and deuterons on a gold target at energies $E_{cm}=200\cdot
A$~GeV.  The d-Au interactions \cite{dumitru,arsene} are very close to p-Au
collisions and therefore interesting for air shower physics.

The OPERA \cite{opera,mauri} neutrino oscillation experiment has been designed
to prove the appearance of $\nu_\tau$ in a nearly pure $\nu_\mu$ beam produced
at CERN and detected in the underground Gran Sasso Laboratory, 730 km away from
the source. In OPERA, $\tau$ leptons resulting from the interaction of
$\nu_\tau$ are produced in target units made of nuclear emulsion films
interleaved with lead plates.  Data taking has started in Summer 2008.

\section{Solar Particles}\label{solsec}

\begin{figure}\centering
 \epsfig{file=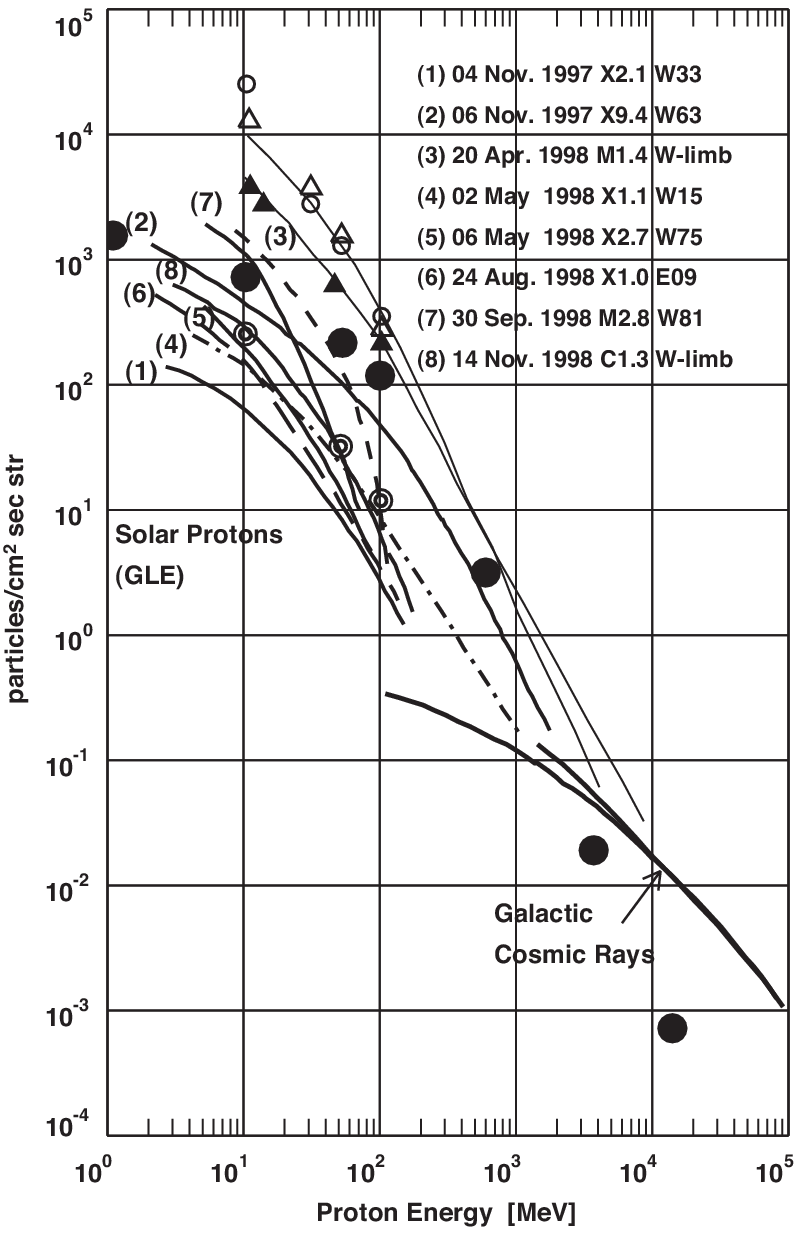, width=\columnwidth}
 \Caption{Integral flux of solar protons. Black dots represent the 2001 Easter
        event, for details see \cite{muraki,muraki-easter}.} 
 \label{muraki}
\end{figure}

The dynamical motion of magnetic loops at the surface of the Sun is the origin
of solar flares \cite{muraki}. The flares are caused by reconnection of
magnetic field lines.  This causes the formation of a plasma jet. Plasma
particles in the magnetic loop are accelerated by collisions with the
high-speed plasma jet from the top of the magnetic loop. The acceleration
mechanism is consistent with a shock acceleration picture.

The integral flux of solar protons is shown in \fref{muraki} together with the
flux of galactic cosmic rays. Depending on the state of the Sun galactic
particles start to dominate the overall spectrum at energies exceeding about
100~MeV to 1~GeV.  Typically, cosmic particles with energies up to about
100~MeV are assumed to originate from the sun. However, recent measurements
indicate that during solar bursts particles can be accelerated to significantly
higher energies, exceeding 50~GeV.

A particularly strong solar event occurred on Easter day 2001 (April 15), which
has been observed and discussed by several groups
\cite{muraki-easter,sep-shea,sep-bieber,sep-tylka,sep-vashenyukasr,braunburst,sep-dandrea}.
The event has been registered with the Karlsruhe Muon Telescope at energies
around $10-20$~GeV \cite{braunburst}.  The GRAND experiment \cite{sep-dandrea}
has detected solar energetic particles exceeding 56~GeV.  The energy spectrum
of the Easter event is presented in \fref{muraki} as filled dots.  Above
200~MeV the neutron monitor data and the GRAND measurements are described by a
single power law with a spectral index $\gamma=-3.75\pm0.15$ \cite{muraki}.
This event shows clear evidence that particles with energies exceeding 50~GeV
are accelerated at the Sun. It should be noticed that applying the Hillas
formula \cite{hillasdiagram} to the Sun yields maximum energies of about
10~TeV.

\section{Direct Measurements of Cosmic Rays}\label{directsec}

\begin{figure}
 \epsfig{file=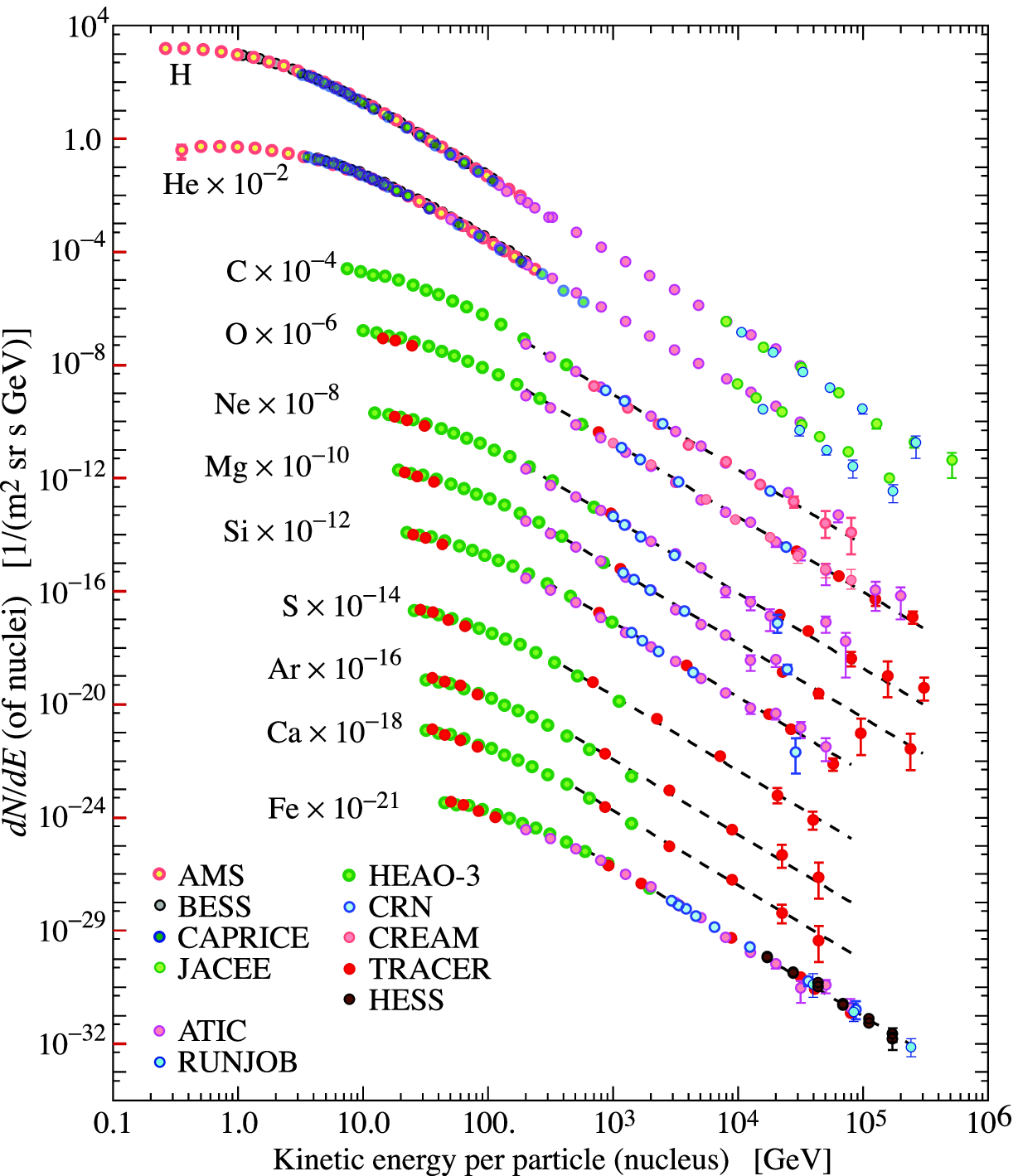, width=\columnwidth}
 \Caption{Energy spectra of main elements in cosmic rays \cite{wefel} and
          \cite{pdg08} (pg. 254).}
 \label{dirspec}
\end{figure}

With the latest generation of experiments, direct measurements of cosmic rays
above the atmosphere now extend to energies around $10^{14}$~eV with
single-element resolution \cite{wefel}. Spectra for main elements in cosmic
rays are summarized in \fref{dirspec}. This compilation covers an enormous
range of 7 orders of magnitude in (total particle) energy.  To reduce overlap
the spectra for individual elements are shifted in vertical direction as
indicated.  Data are shown from the magnet spectrometers AMS \cite{amsp}, BESS
\cite{bess}, and CAPRICE \cite{caprice98}; from the calorimeters ATIC
\cite{atic06} and CREAM \cite{maestro,creamexp}; from the transition radiation
detectors CRN \cite{crn} and TRACER \cite{gahbauer,tracer05,muellermerida};
from the \Cerenkov detector hodoscope HEAO-3 \cite{heao3}; as well as from the
emulsion chamber experiments JACEE \cite{jaceephe} and RUNJOB \cite{runjob05}.

To extend the measurements of energy spectra up to the knee ($>10^{15}$~eV) the
Advanced Cosmic Ray Composition Experiment for the Space Station (ACCESS) has
been proposed \cite{access}. NASA initiated a balloon program in support for
ACCESS: the transition radiation detector TRACER, the calorimeter ATIC, and
CREAM, combining both techniques. Unfortunately, NASA did not select ACCESS
for flight neither on the ISS nor as a free-flyer, leaving the balloon payloads
as the source of new results. All three experiments are part of the Long
Duration Balloon program of NASA \cite{ldb}, providing circumpolar flights from
Mc Murdo, Antarctica. The present record holder in floating time is the CREAM
experiment with balloon flights in the seasons 2004/5, 2005/6, and 2007/8 with
flight durations of 42, 28, and 29 days, respectively.  As can be inferred from
\fref{dirspec} the three experiments deliver results at the highest energies.

The abundances of elements heavier than iron are measured with TIGER
\cite{tiger}. 

PAMELA, a new space based experiment has been launched in June 2006 and first
data have been published \cite{adriani}. Main components are a magnet
spectrometer, a time-of-flight system, an electromagnetic calorimeter, and a
neutron detector. PAMELA is expected to yield precise information on electrons,
positrons, cosmic-ray nuclei, and anti-nuclei up to energies of several
100~GeV.

\begin{figure}\centering
 \epsfig{file=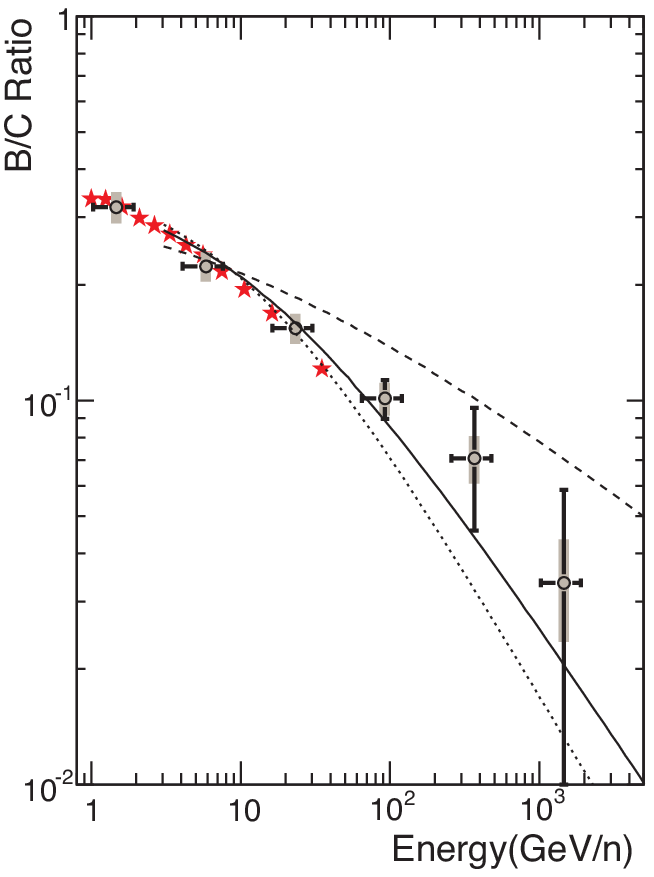, width=0.90\columnwidth}
 \Caption{Boron-to-carbon ratio in cosmic rays as measured by the CREAM balloon
experiment (circles) and the HEAO-3-C2 space experiment (asterisks)
\cite{creambc}. The thin lines represent statistical uncertainties, the grey
bands indicate systematic uncertainties.}
 \label{creamBC}
\end{figure}

To distinguish between different propagation models of cosmic rays in the
Galaxy it is of interest to measure the ratio of secondary to primary cosmic
rays, e.g.\ the boron-to-carbon ratio.  \footnote{Attention should be payed to
not confuse {\sl secondary cosmic rays}, produced during the propagation
process in the Galaxy, with {\sl secondary particles}, produced in air showers
inside the atmosphere.} Measurements indicate that propagation in the Galaxy is
energy dependent with the mean path length $\lambda$ traversed decreasing with
increasing energy $\lambda=\lambda_0(R/R_0)^\delta$, with the rigidity $R=E/z$,
where $E$ is the energy and $z$ the charge of the particle.  Current data, up
to about 100 GeV/n, indicate a fall-off in energy with an exponent of
$\delta\approx-0.5$ to $-0.6$.  Theories suggest that at high energy the
exponent should approach $-1/3$ for a Kolmogorov turbulence spectrum. It is
therefore of great interest to extend such measurements to energies as high as
possible.  

Recent results obtained by the CREAM experiment are depicted in \fref{creamBC},
they extend measurements of the B/C ratio up to $10^3$~GeV/n \cite{creambc}.
The lines represent model calculations with values for $\delta=-0.33$ (dashed
line), $-0.6$ (solid line), and $-0.7$ (dotted line). At high energies the data
are compatible with $\delta\approx -0.6$ to $-0.5$.  It is of interest to point
out that the grey bars indicate the systematic uncertainties of the
measurements.  The dominant source of systematic uncertainties at high energy
are uncertainties in the cross sections for producing secondaries by
charge-changing interactions in the atmosphere above the instrument. Because
there is a significant decrease in the interstellar path length with energy,
the amount of boron, for example, at high energy is small --- making the impact
of uncertainties in the atmospheric boron contribution significant above
$\approx100$~GeV/n.  At these energies the contribution from charge-changing
interactions in the atmosphere is similar in size to the total production of
boron during propagation through the Galaxy.  This emphasizes the importance to
measure such cross sections with high precision at accelerators.

One of the hottest topics recently discussed is the discovery of anomalies in
the fluxes of electrons and positrons.  The ATIC collaboration finds an excess
of electrons at energies of $300-800$~GeV \cite{aticnature}.  The PAMELA group
measured an excess in the positron-to-electron fraction $e^+/(e^-+e^+)$ at
energies between 10 and 100~GeV \cite{Adriani:2008zr}.  A high precision
measurement of the electron spectrum in the energy range from 20~GeV to 1~TeV
has been obtained with the Large Area Telescope (LAT) on board the Fermi
satellite \cite{Abdo:2009zk}.  The recent data stimulated numerous attempts for
a theoretical interpretation of the observed structures (a few hundreds of
articles on arXiv).  Common explanations include modifications of the diffuse
background model due to local sources, contributions of local astrophysical
sources such as pulsars, reacceleration at supernova remnants, or dark matter
annihilation.

\section{TeV Gamma-Ray Astronomy}\label{gammasec}

Many new results from the TeV gamma-ray telescopes H.E.S.S. \cite{punch} and
MAGIC \cite{majumdar} have been presented. Here we will focus on results which
are in particular related to cosmic-ray physics.  The measurements of TeV gamma
ray emission from supernova remnants strongly supports the theory of Fermi
acceleration of hadronic cosmic rays in these objects, see e.g.\
\cite{Voelk:2008fw,Berezhko:2007gh}. They provide evidence that (at least a
large fraction of) galactic cosmic rays are accelerated in the vicinity of
supernova remnants.

The H.E.S.S. \Cerenkov telescope system derived for the first time an energy
spectrum measuring direct \Cerenkov light \cite{Aharonian:2007zja}.  This light
is emitted by the primary nucleus in the atmosphere before its first
interaction, i.e.\ before the air shower begins \cite{directc}.  The energy
spectrum of iron nuclei has been reconstructed in the energy range between 10
and 100~TeV. The new method is very valuable to bridge the gap between direct
measurements and classical air shower measurements, using particle detector
arrays.

\begin{figure}
 \epsfig{file=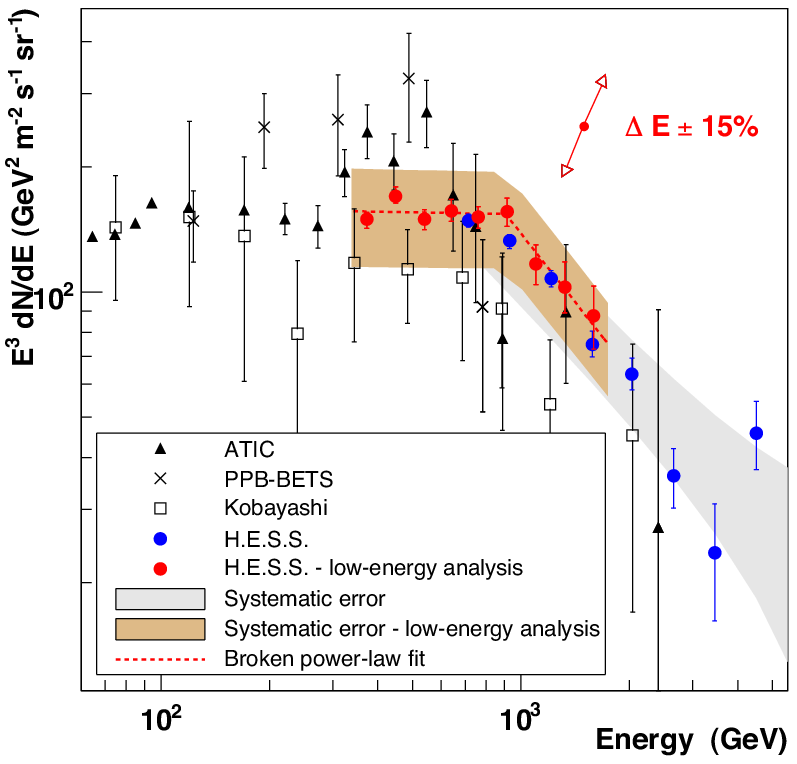, width=\columnwidth}
 \Caption{Energy spectrum of primary cosmic-ray electrons $E^3 dN/dE$ as
          observed by several experiments \cite{Aharonian:2009ah}.}
 \label{hesselectr}
\end{figure}

Also for the first time the primary cosmic-ray electron spectrum has been
measured with a ground based detector, the H.E.S.S. experiment
\cite{Collaboration:2008aaa,Aharonian:2009ah}. An effective gamma-hadron
separation algorithm is applied to infer the electron spectrum between 0.3~TeV
and about 10~TeV. The result is depicted in \fref{hesselectr}. The H.E.S.S.
data are compared to measurements with instruments above the atmosphere. A
depression of the flux at energies exceeding 1~TeV can be inferred. It is
interesting to note that the H.E.S.S. data are compatible with the Fermi
measurements at energies around 0.5~TeV, see also last paragraph of
\sref{directsec}.

\section{Neutrino Astronomy}\label{neutrinosec}

In addition to information from charged particles and gamma rays, neutrino
astronomy is expected to yield complementary information to reveal the origin
of ultra high-energy cosmic rays. 

The IceCube neutrino telescope \cite{berghaus,DeYoung:2009er} is presently
under construction at the geographical South Pole. It is the successor of the
AMANDA experiment, operated till 2009. It is a km$^3$ detector, comprising 80
strings of 4800 optical modules in total and $80\times2$ surface detectors
(IceTop) \cite{stanev}. At the time of the conference 80 surface detectors and
40 strings have been deployed.  The surface detectors are ice \Cerenkov
detectors, each equipped with two optical modules.

\begin{figure}\centering
 \epsfig{file=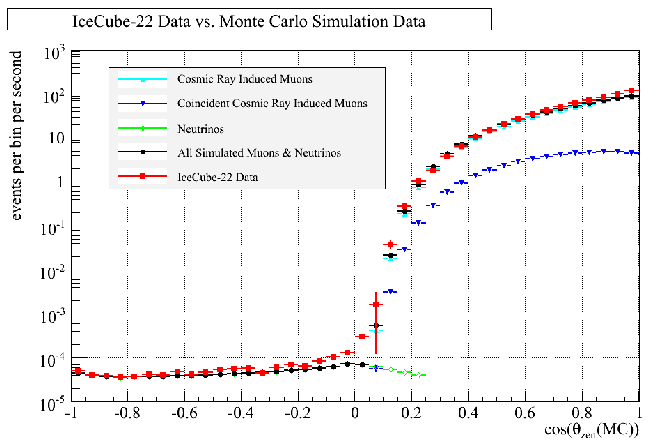, width=\columnwidth}
 \Caption{All-sky muon flux as measured with IceCube (22 strings)
  \cite{berghaus}.}
 \label{icecubemuons}
\end{figure}

Data taking and analysis has started with the parts of the experiment already
installed. As an example of recent results, the all-sky muon flux as measured
with 22 strings is presented in \fref{icecubemuons} \cite{berghaus}.  Using a
combination of minimum bias and muon-filtered data, the performance of IceCube
has been verified by comparing the measured muon track zenith angle
distribution to a Monte Carlo simulation.  For the first time the muon flux was
measured over the entire sky using an uniform set of quality parameters.  Data
and simulation generally agree to better than 20\%.

Complementary to the South Pole, neutrino telescopes are under construction in
the Mediterranean  See.  The installation of ANTARES \cite{kouchner} has just been
completed. It comprises 12 strings, each equipped with 75 photo multipliers
arranged in 25 "storeys". Data taking has started and first analyses are under
way.

The activities of the neutrino telescopes ANTARES, NEMO, and NESTOR are
bundled in the KM3NET \cite{dornic} consortium to build a km$^3$ neutrino
detector in the Mediterranean  See. At present, various design studies are
conducted, construction of the detector is expected to start in 2011
\cite{km3net-cdr}.

The Pierre Auger Observatory serves also as detector for high-energy neutrinos
\cite{gora,Abraham:2007rj}. The analysis of horizontal air showers yields
limits on the neutrino flux in the energy range around $10^{17}$ to
$10^{19}$~eV which already at present severely constrain exotic models for the
origin of ultra high-energy cosmic rays. The upper limits are only about one
order of magnitude above the flux expected from neutrinos produced during the
propagation of cosmic rays (cosmogenic neutrinos, GZK neutrinos).

\section{High-Energy Interactions in the Atmosphere}\label{wwsec}

Air shower data can be used to deduce properties of high-energy hadronic
interactions in the atmosphere. This is of particular interest in energy
regimes and kinematical regions beyond the capabilities of present-day
accelerators.

Frequently, high-energy muons are used to study details of the shower
development \cite{kokoulin,petrukhin}. Measurement of their production height
through triangulation with a tracking detector \cite{doll} allows insight into
properties of individual hadronic interactions \cite{zabierowski}.
The OPERA experiment has obtained a new measurement of the muon charge ratio
\cite{mauri}.

The KASCADE experiment, measuring simultaneously the hadronic, the
electromagnetic, and the muonic shower components is in particular sensitive to
details of hadronic interaction models. The data are used to check the
consistency of the predictions of hadronic interaction models used in air
shower simulations with measured data
\cite{wwtestjpg,jensjpg,epostest,jrhepos}. It could be shown that the recent
model EPOS, version 1.6 is not compatible with air shower data. This stimulated
the development of a new version 1.9, which is presently under investigation.

\begin{figure}
 \epsfig{file=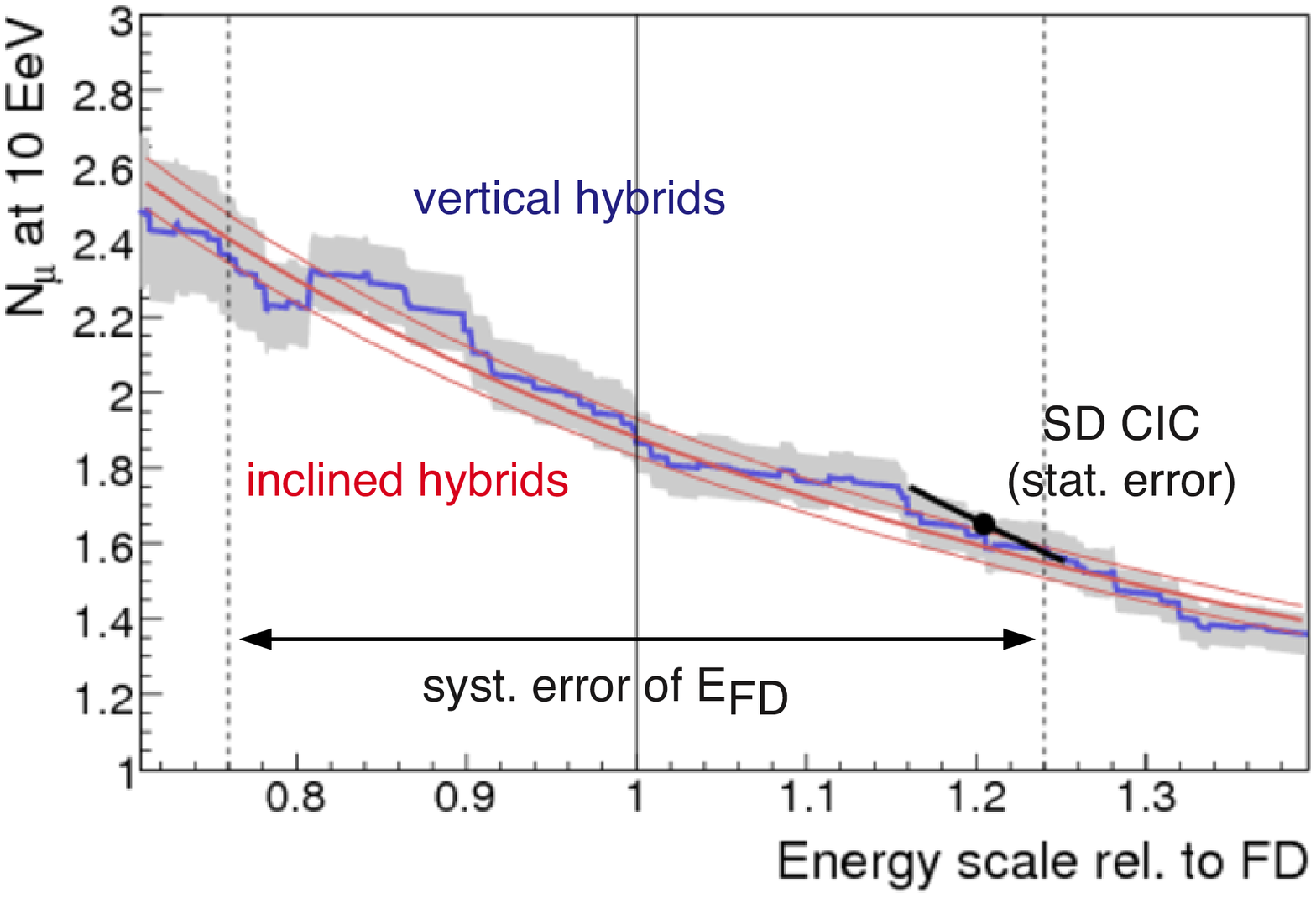, width=\columnwidth}
 \Caption{Comparison of the results on the number of muons $N_\mu$ at 10~EeV
	  from different methods \cite{Schmidt:2009ge}. Results for vertical
	  and inclined hybrid events are shown as well as results applying the
          constant intensity method.}
 \label{augermu}
\end{figure}

The muon content at ground level in the data from the Pierre Auger Observatory
has been analyzed in detail at energies around 10~EeV
\cite{Schmidt:2009ge,schmidt}. Different sets of hybrid events were used, i.e.\
the showers have been observed simultaneously with the fluorescence telescopes
and the surface detectors. And a constant intensity cut method has been applied
to data from the surface detectors.  The number of muons is shown as function
of the energy relative to the energy scale of the fluorescence detector in
\fref{augermu}.  The investigations reveal that the hadronic interaction models
predict not enough muons at the highest energies.

\subsection{Proton-air cross section}

\begin{figure}
 \epsfig{file=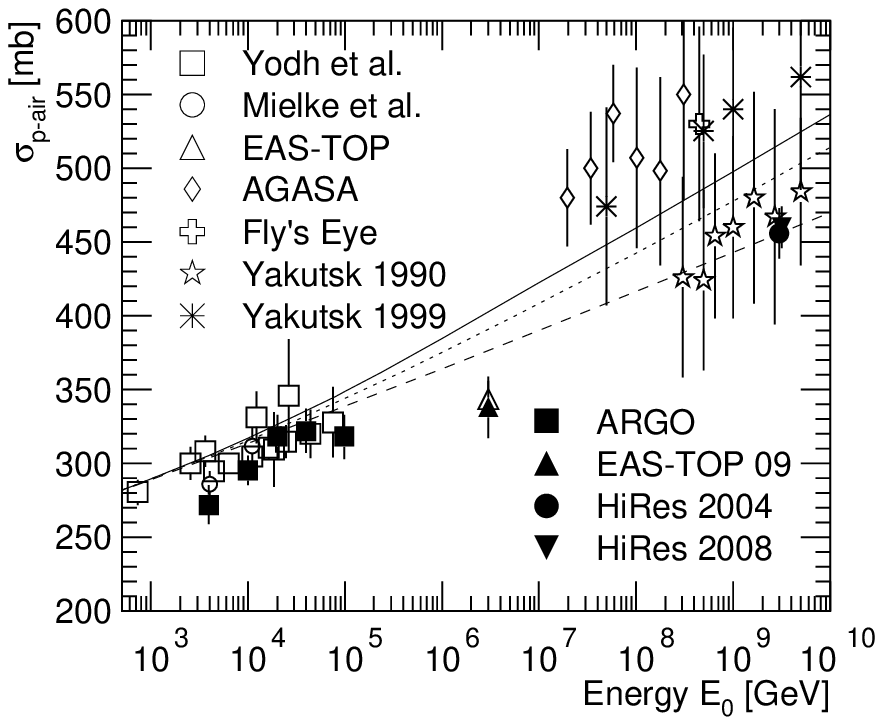, width=\columnwidth}
 \Caption{Inelastic proton-air cross section as function of primary energy.
For references of the experimental values see text.  The lines
          represent calculated cross-sections for three different versions of
          the model QGSJET~01, model~1 (original, ---),
          model~2~($\cdot\cdot\cdot$), and model~3 (- - -), see \rref{wq}.}
 \label{pair}
\end{figure}

The inelastic proton-air cross section has been determined from air shower
data. Three techniques can be distinguished to determine the proton-air cross
section: investigating unaccompanied hadrons by Yodh \etal \cite{yodh72,yodh83}
and Mielke \etal \cite{mielkesh}; the $N_e/N_\mu$ technique, carefully
selecting proton like events as AGASA \cite{hara,honda}, ARGO
\cite{Aielli:2009}, EAS-TOP 1999 \cite{eastopwq99} and 2009
\cite{eastopwqprd,trinchero}; as well as analyzing the tail of the measured
\Xmax distributions by Fly's Eye \cite{flyseyewq84,flyseyewq85}, HiRes 2004
\cite{belovisvhecri} and 2009 \cite{belovmerida}, and Yakutsk \cite{dyakonov}.
The experimental values are compiled in \fref{pair} and compared to predictions
of the hadronic interaction model QGSJET~01 and a modification, see \rref{wq}.
It is interesting to realize that all recent measurements are at the lower edge
of the range of measured values.

A thorough systematic study of the effects of the proton-air cross section on
air shower observables has been presented by Ulrich \cite{ulrich}.  The
dependence of the depth of the shower maximum \Xmax, as well as the number of
electrons and muons on the inelastic proton-air cross section has been
investigated.

The LHC forward detectors, see \sref{accelsec} will significantly contribute to
a better understanding of the proton-air cross section at high energies in the
near future.

\subsection{Radio Detection of Air Showers}
Presently, a very promising technique to detect air showers is revitalized: the
measurement of radio emission from air showers \cite{Falcke:2008qk}.  Two
experiments are taking data in the energy range around $10^{17}$~eV: CODALEMA
\cite{belletoile,Ardouin:2009zp} and LOPES \cite{haungs,radionature}.  They
register radio frequencies in the range of several tens of MHz.  Results of
both experiments show clear evidence for a geomagnetic origin of the measured
emission in air showers.  Most likely, the radio emission is synchrotron
radiation generated by electrons and positrons (most of them with energies
slightly lower than the critical energy $E_{crit}\approx84$~MeV) in the
magnetic field of the Earth.

A new digital radio telescope LOFAR \cite{jrhlofar} is presently under
construction in the Netherlands and in Europe. With a very high density of
antennas (several hundreds of antennas within 2~km diameter in the core) this
instrument will be very valuable to study the radio emission in detail, e.g.\
through precise measurements of the curvature of the shower front.

\begin{figure}\centering
 \epsfig{file=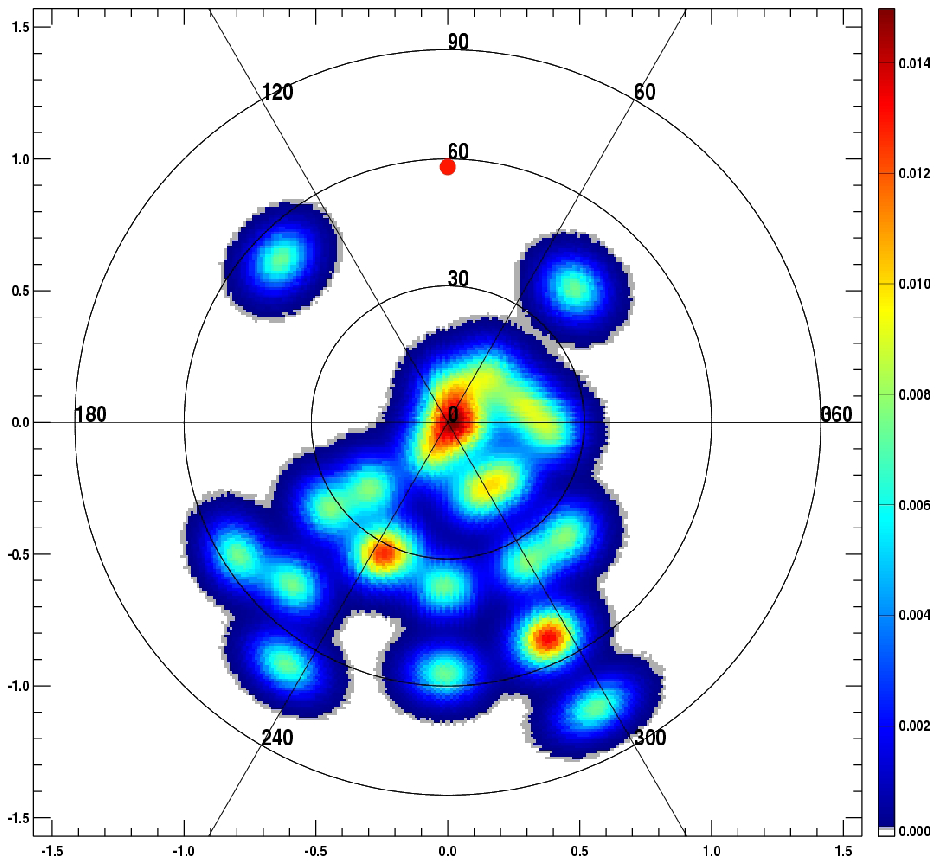, width=\columnwidth}
 \Caption{Arrival direction of cosmic rays registered with radio antennas at
  the Pierre Auger Observatory \cite{belletoile}.}
 \label{radiosky}
\end{figure}

Radio emission from air showers is also studied at the Pierre Auger Observatory
in Argentina \cite{vdbergmerida}.  Prototype studies are conducted with set-ups
evolving from the CODALEMA and LOPES experiments.  As an example, the arrival
direction of air showers registered with radio antennas is depicted in
\fref{radiosky}. The arrival directions are clearly non-isotropic, 20 out of 25
events arrive from the South.  The maximum number of events measured in
East-West polarization direction is registered at arrival directions with an
angle of about 90\deg to the direction of the Earth magnetic field.  Similar
investigations in the northern hemisphere with CODALEMA and LOPES reveal the
same effect. This provides clear evidence for a geomagnetic origin of the
observed emission.  Main focus of the future activities is the construction of
the Auger Engineering Radio Array (AERA) comprising $\approx150$ antennas on an
area of about 20~km$^2$ to measure air showers in the energy range from about
$10^{17}$ to $10^{19}$~eV \cite{vandenberglodz}.

\section{Indirect Measurements of Cosmic Rays} \label{eassec}

\begin{figure*}\centering
 \epsfig{file=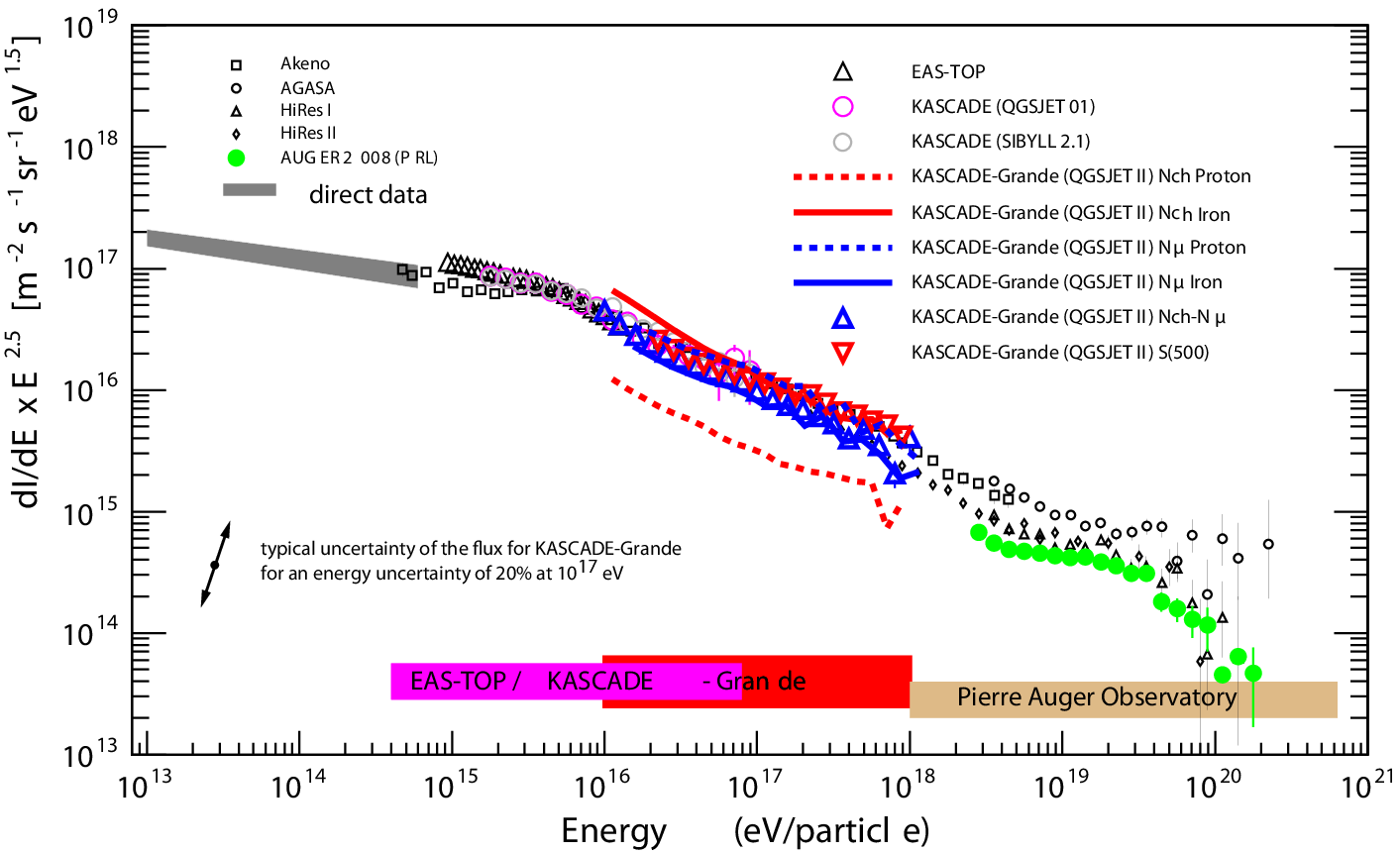, width=0.8\textwidth}
 \Caption{All-particle energy spectrum of cosmic rays. Spectra obtained with
	  different methods applied to KASCADE-Grande data are compared to
          results of other experiments  \cite{kascadeicrc09}.}
 \label{kgall}
\end{figure*}

The anisotropy of cosmic rays in the energy region from $10^{13}$ to
$10^{14}$~eV has been studied by the Baksan group \cite{alexeenko}. An
amplitude of a few times $10^{-4}$ has been observed.

In the energy region of the knee in the spectrum ($\approx10^{15}$~eV) the
Gamma experiment derived an all-particle energy spectrum \cite{martisov}.
The Tibet group reconstructed spectra for primary protons and helium nuclei as
well as the all-particle spectrum
\cite{huang,Amenomori:2008jb,Amenomori:2005nx}.  Spectra for five elemental
groups (p, He, CNO, Al, Fe) have been reconstructed with data from the GRAPES
experiment \cite{gupta,grapes05}.
KASCADE presented an update on the unfolding of the spectra for groups of elements, different zenith angle intervals have been analyzed and different hadronic interaction models have been used to interpret the data --- the findings 
confirm earlier results \cite{haungs,ulrichapp,Apel:2008cd}.
The spectra obtained by the different groups fit well into the world
data sample and contribute to a consistent picture in the energy region of the
knee, see e.g.\ \rref{cospar06} and \cite{behreview} for a comparison of the
data.
The Tibet group presented ambitious plans to extend their installation. The
planned set-up comprises a scintillator array for the electromagnetic component
as well as muon detectors: an array of shielded scintillation counters and
underground water \Cerenkov detectors.

In current astrophysical models a transition from a galactic to an
extra-galactic origin of cosmic rays is expected at energies around $10^{17}$ to
$10^{18}$~eV \cite{behreview}.
A precise measurement of the mass composition in this energy region will be
important to distinguish between different astrophysical scenarios.

Several methods are applied to data from the KASCADE-Grande experiment to
reconstruct the all-particle spectrum between $10^{16}$ and $10^{18}$~eV
\cite{haungs,kascadeicrc09}: the constant intensity cut method is applied to
measurements of charged particles and muons; the density of charged particles
measured at a distance of 500~m to the shower axis S(500) is used; and
an unfolding algorithm is applied to electron and muon number data to obtain spectra for groups of elements.
The all-particle spectra obtained with the different methods are shown in
\fref{kgall} and are compared to results of other experiments.
The KASCADE-Grande data smoothly continue the all-particle spectrum obtained by
KASCADE.
Detailed studies of the mass composition are under way.

The 1~km$^2$ IceTop array \cite{stanev} at the South Pole is half completed,
see also \sref{neutrinosec}. Data taking has started with the existing parts.
First analyses are under way, investigating the lateral distribution of air
showers and shower size spectra.

The origin of the highest energy cosmic rays is presently explored with several
instruments: the Pierre Auger Observatory is taking data on its southern site
\cite{engel}, also the Telescope Array \cite{tameda} is fully operational,
while the HiRes experiment \cite{sokolsky} has stopped data taking.  A
suppression of the flux has been observed at energies exceeding
$4\cdot10^{19}$~eV by the HiRes experiment \cite{Abbasi:2007sv} and the Pierre
Auger Observatory \cite{Abraham:2008ru}.
One of the most exciting results of the last years is the discovery of the
Auger Observatory that the arrival directions of the highest energy cosmic rays
are non-isotropic and they are correlated with the position of the
supergalactic plane on the sky \cite{agnscience,Abraham:2007si}.

\begin{figure}
 \epsfig{file=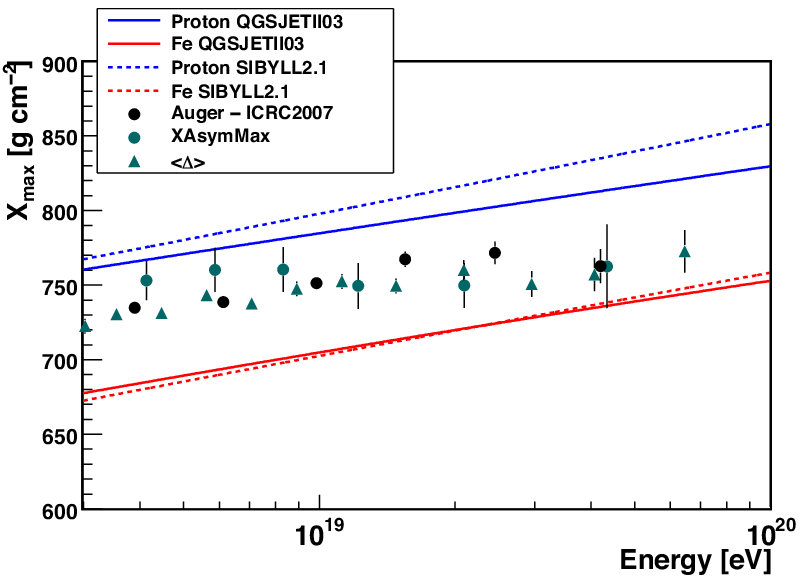, width=\columnwidth}
 \Caption{Average depth of the shower maximum \Xmax as function of energy,
	  according to different methods applied to data from the Pierre Auger
	  Observatory. The measured values are compared to predictions for
	  primary protons and iron nuclei according to two hadronic interaction
          models as indicated  \cite{Abraham:2009ds}.}
 \label{augerxmx}
\end{figure}

Different methods are applied to reconstruct the mass composition from Auger
data, i.e.\ the average depth of the shower maximum \Xmax
\cite{Abraham:2009ds,wahlberg}: \Xmax is observed directly with the
fluorescence telescopes and it is derived from data of the surface detectors
investigating the rise time of the signals in the water \Cerenkov detectors as
well as evaluating the azimuthal asymmetry of the arrival time distributions.
The values obtained are compiled in \fref{augerxmx}. A consistent trend towards
a heavier mass with increasing energy can be recognized.

It is an interesting challenge to combine the latest observations in an
astrophysical scenario: the depression in the energy spectrum, the observed
anisotropy of the arrival directions, and the development of the mass
composition as function of energy. A correlation of the arrival directions with
nearby ($<100$~Mpc) astronomical objects requires particles with a small charge
to have only small deflections in the galactic (and extra-galactic) magnetic
fields. A light composition would also be expected if the observed depression
is the GZK effect \cite{gzkgreisen,cronin-gzk}. On the other hand, the data
exhibit an increase of the average mass as function of energy.  It is also
possible that the depression in the spectrum is due to the maximum energy
attained during the acceleration process. In this case one would expect a
rigidity dependent cut-off behavior of individual elements (similar to the one
observed at the knee, but at much higher energies) and one would obtain an
increase of the mean mass as function of energy.

At present, the Auger Collaboration is preparing the northern site of the
Observatory with an area covering at least 10000~km$^2$ \cite{anorthmerida}.
This will be an important and mandatory step towards astronomy with charged
particles on the full sky.

New experiments are under preparation to observe the fluorescence light of
extensive air showers from above the atmosphere. The TUS mission is planned for
2010 with the aim to measure about 10 events/year in the $10^{20}$~eV energy
region \cite{tkatchev}.  JEM-EUSO is expected to be attached to the Japanese
module of the ISS in 2013 to measure cosmic rays with energies exceeding
$10^{19}$~eV \cite{inoue}.

\section{Conclusions and Outlook}

At the conference a wealth of new data has been presented. They improve our actual knowledge about the origin of high-energy cosmic rays.
In the next years many new significant contributions are expected to reveal the
origin of high-energy cosmic particles.

The LHC will provide new data on high-energy interactions which will be
extremely useful for the interpretation of air shower data.

The atmospheric \Cerenkov telescopes H.E.S.S., MAGIC, and VERITAS are taking
data and are upgrading their detectors. TeV gamma-ray astronomy has already
made significant contributions to the understanding of the sources and the
acceleration mechanisms for (charged) cosmic rays. The coming years are
expected to lay open more secrets of galactic and extra-galactic sources.

New balloon experiments are under preparation. They will improve our
understanding of the propagation of cosmic rays in the Galaxy.

The GRAPES and Tibet groups are currently upgrading their detectors and will
provide new data on the composition of cosmic rays in the energy region of the
knee in the spectrum ($\approx10^{15}$~eV).

Several installations are devoted to the energy region of the transition from
galactic to extra-galactic cosmic rays ($10^{17}-10^{18}$~eV): KASCADE-Grande,
IceCube/IceTop, the low-energy extensions of the Pierre Auger Observatory
(HEAT, AMIGA, AERA), as well as the low-energy extension of the Telescope Array
(TALE). A precise measurement of the mass composition in this energy region
will be crucial to discriminate between different astrophysical scenarios.

The origin of extra-galactic cosmic rays is investigated with set-ups operating
at the highest energies: the Telescope Array is fully operational, the southern
site of Pierre Auger Observatory is fully operational, the northern site is
under preparation, JEM-EUSO is progressing fast. 

Also the neutrino telescopes are expected to deliver significant data, ANTARES
is fully operational and KM3NET is on good track, IceCube is expected to be
completed in 2011. 

In summary, we are facing exciting times with big experimental challenges, they
will offer unique opportunities and will provide new insights into the
(astro)physics of the Universe in the next decade.

\section{Acknowledgments}
It was a pleasure to summarize the experimental results of this meeting. 
I would like to thank the organizers for offering such an interesting
scientific program.
The author is grateful to valuable discussions with his colleagues from the
Pierre Auger Observatory, as well as from the KASCADE-Grande, LOPES, LOFAR, and
TRACER experiments.


\end{document}